\documentclass{PoS}

\PoS{PoS(LAT2005)038}

\usepackage{epsfig}

\title{Automated Methods in Chiral Perturbation Theory on the Lattice }

\ShortTitle{Automated Methods in Chiral Perturbation Theory on the Lattice }

\author{Bu\=gra Borasoy$^a$, \speaker{Georg M. von Hippel}$^b$,
        Hermann Krebs$^a$ and Randy Lewis$^b$\\ \\
        \llap{$^a$}Helmholtz-Institut f\"ur Strahlen- und Kernphysik (Theorie),
	           Universit\"at Bonn, Bonn, Germany\\
        \llap{$^b$}Department of Physics, University of Regina, Regina,
	           Saskatchewan, Canada\\

        E-mail: \email{vonhippg@uregina.ca}}

\abstract{
We present a method to automatically derive the Feynman rules for
mesonic chiral perturbation theory with a lattice regulator. The
Feynman rules can be output both in a human-readable format and in
a form suitable for an automated numerical evaluation of lattice
Feynman diagrams. The automated method significantly simplifies
working with improved or extended actions. Some applications to the
study of finite-volume effects will be presented.
}

\FullConference{XXIIIrd International Symposium on Lattice Field Theory\\

                 25-30 July 2005\\

                 Trinity College, Dublin, Ireland}

\begin{document}

\section{ Introduction }

\subsection{ Chiral Perturbation Theory on the Lattice }

One of the most important uses of Chiral Perturbation Theory ($\chi$PT)
is the extrapolation of results from Lattice QCD simulations.
Finite volume effects, discretization errors and quark mass
extrapolations can all be addressed using $\chi$PT.
In this context, $\chi$PT can be regulated with either a continuum formalism
\cite{baer} or a lattice formalism \cite{lewis}.
Though the lattice approach typically leads to more complicated algebra,
it has the advantage of being directly amenable to numerical methods.
A recent study of volume dependences in 1-loop lattice $\chi$PT demonstrates
how simple numerical methods are able to replace extensive algebra
\cite{borasoy}. Besides its relevance to QCD simulations, lattice
$\chi$PT is vital for nuclear simulations with chiral effective
theories \cite{lee}.

In lattice regularized $\chi$PT , the meson fields
live on the lattice sites, and gauge fields as usual on the
links. Derivatives in the Lagrangian are replaced by appropriately
covariantised finite differences. Lattice Chiral Perturbation Theory
has been used to compute baryon magnetic moments \cite{borasoy2} and
to simulate neutron matter \cite{lee}.

Since Feynman rules become much more complicated on the lattice due to
the appearance of trigonometric functions and the loss of Lorentz
symmetry, deriving them manually becomes an extremely tedious task,
making an automatic procedure desirable.

\subsection{ Automated Generation of Feynman Rules }

Procedures to automatically derive Feynman rules for lattice gauge
theories from their lattice actions have been known since
\cite{luescher}. A new, efficient and adaptable method was presented
in \cite{hart}. The great flexibility of this latter method makes it
possible to adapt it to the case of $\chi$PT on a lattice without much
difficulty by adding support for separate left- and right-vector fields
and extending the SU($N$) algebra to include quark mass matrices. The
structure of a meson $n$-point vertex with mesons of flavours $a_i$
incoming with momenta $p_i$ is
\[
\sum_{\sigma\in\mathcal{S}_n} 
\sum_i \frac{1}{F^{\chi_i}} f_i T_i^{a_{\sigma(1)}\cdots
    a_{\sigma(r_i)}} 
\exp\left(i\sum_j p_{\sigma(j)}\cdot v_{i,j}\right)
\]
where each term can be described by an \emph{entity}
$(f,\chi,T,x,y,\{v_j\})$, with $F$ the usual $\chi$PT parameter (pion
decay constant in chiral limit), $\chi$ the chiral order and $f$ the
amplitude of the term, $T$ a SU($N$) trace, and $x$, $y$ and $v_j$
lattice sites. Entities can be multiplied with the
multiplication rule
\[
(f,\chi,T,x,y,\{v_j\})*(f',\chi',T',x',y',\{v'_{j'}\}) =\]
\vskip-4ex\[
(ff',\chi+\chi',T*T',x,y+y'-x',\{v_j\}\cup\{v'_{j'}+y-x'\})
\]
following from the above representation of the vertex.

Fields can be
expressed in terms of basic entities, and arithmetic operations on
fields can be turned into operations on the entities in their
expansions. Once the field algebra has been implemented, the
perturbative expansion of the action expressed in terms of the fields
into entities will occur automatically.

\begin{figure}
\includegraphics[width=0.55\textwidth]{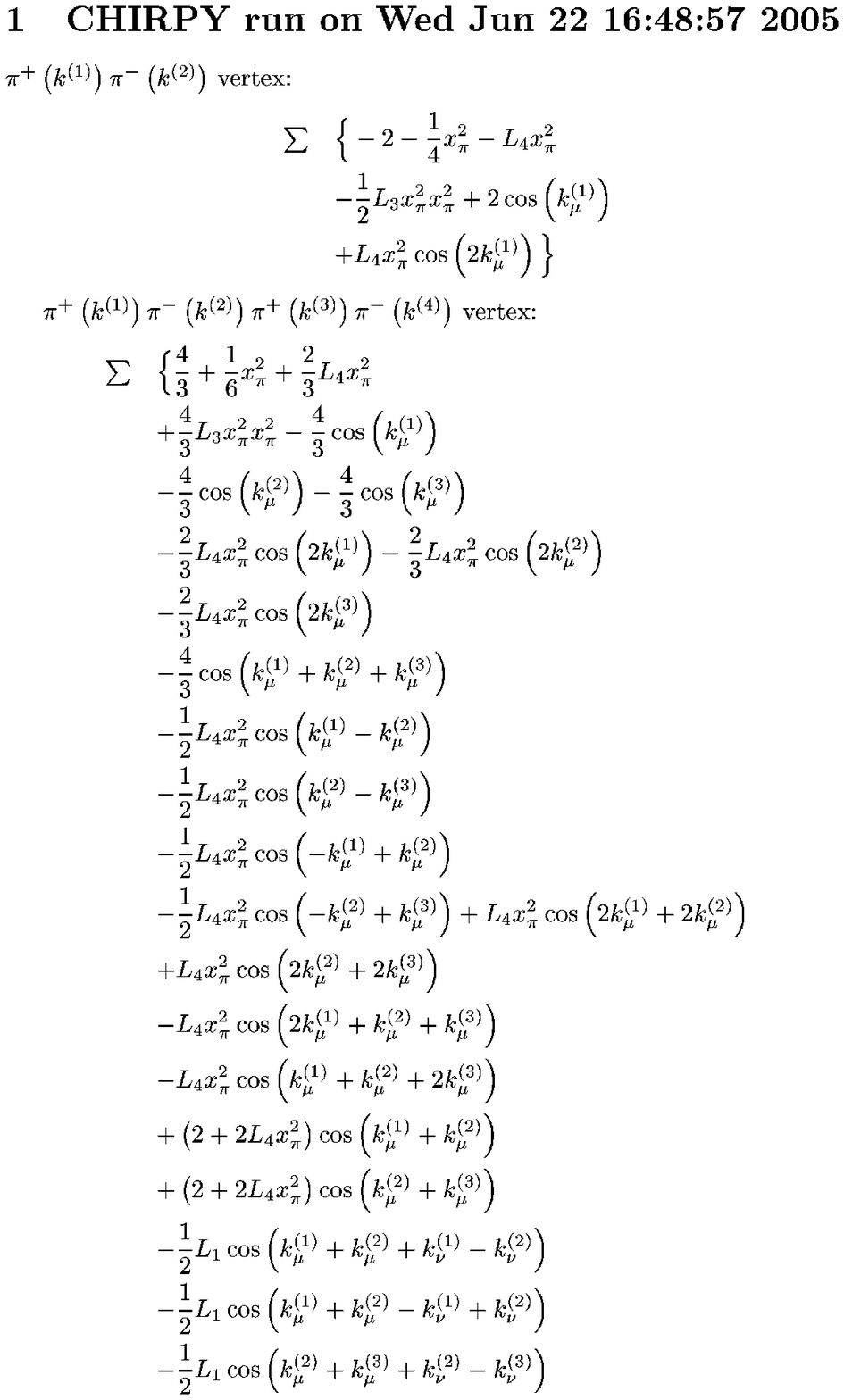} 
\includegraphics[width=0.45\textwidth]{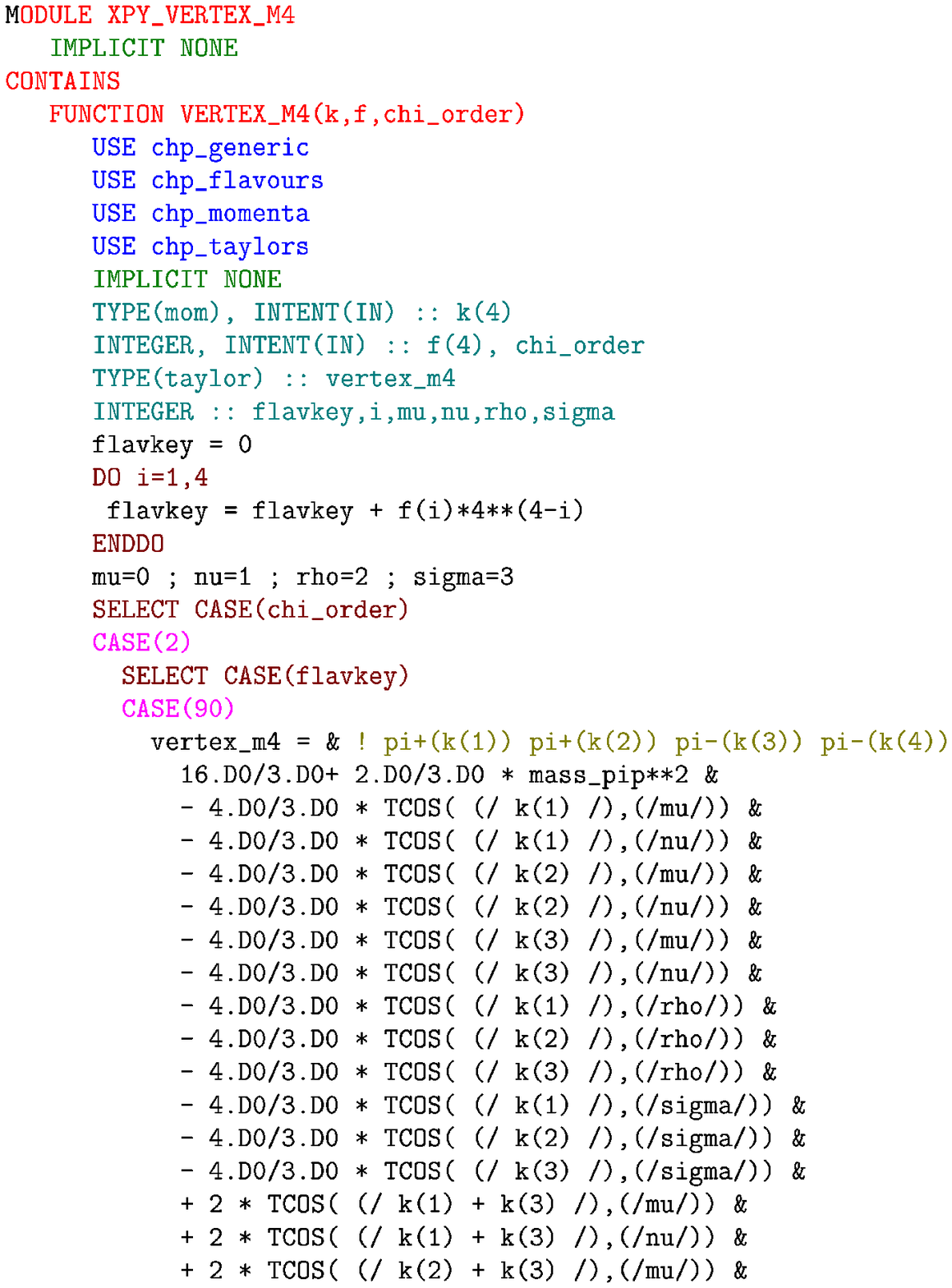}
\caption{ (left) \LaTeX\ output generated by \textsc{chirpy} in human-readable
mode. (right) Fortran output generated by \textsc{chirpy} in
compileable mode. Shown are only the first pages of the multi-page outputs.}
\label{fig:rules}
\end{figure}

The expansion algorithm is wrapped into a \textsc{Python} program
called \textsc{chirpy}, which performs the expansion of an arbitrary
lattice $\chi$PT action given as input and produces the Feynman rules
in a human-readable \LaTeX, compileable Fortran or special
machine-readable format (Fig. \ref{fig:rules}). The relevant flavour,
Lorentz and Taylor \cite{vonhippel} algebra is provided in a companion
Fortran~95 library called \textsc{chirper}. Other output formats
(e.g. for analytical evaluation by a computer algebra program) could
be added easily as required.

\subsection{ Automated Generation of Feynman Diagrams }

There exist a number of well-known packages for the generation of
continuum Feynman diagrams, such as FeynArts \cite{hahn} or Qgraf
\cite{nogueira}. None of these, however, easily support the vertices and
counterterms of arbitrary order that appear in lattice $\chi$PT.

\begin{figure}
\includegraphics[width=0.5\textwidth]{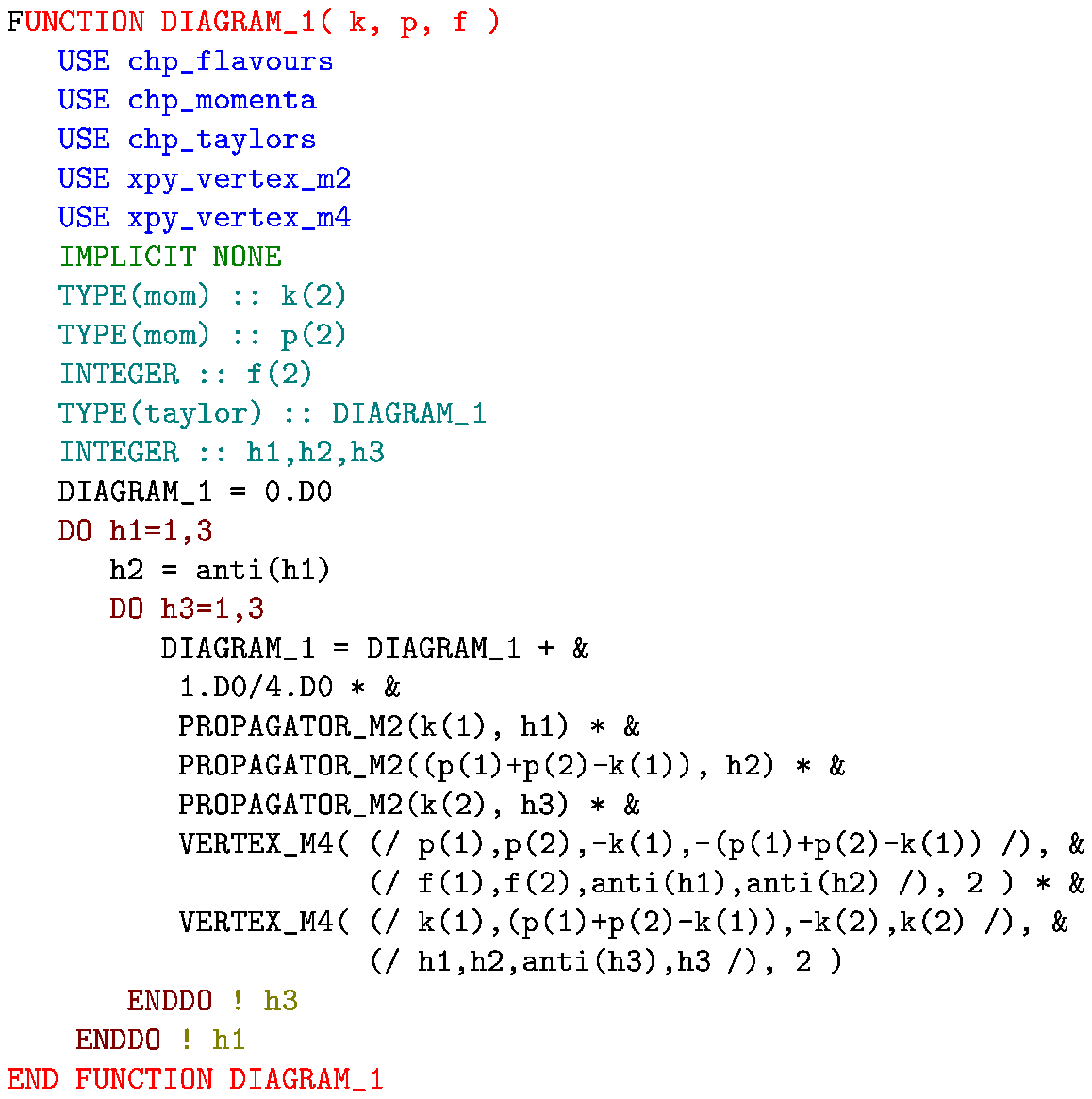}
\includegraphics[width=0.5\textwidth]{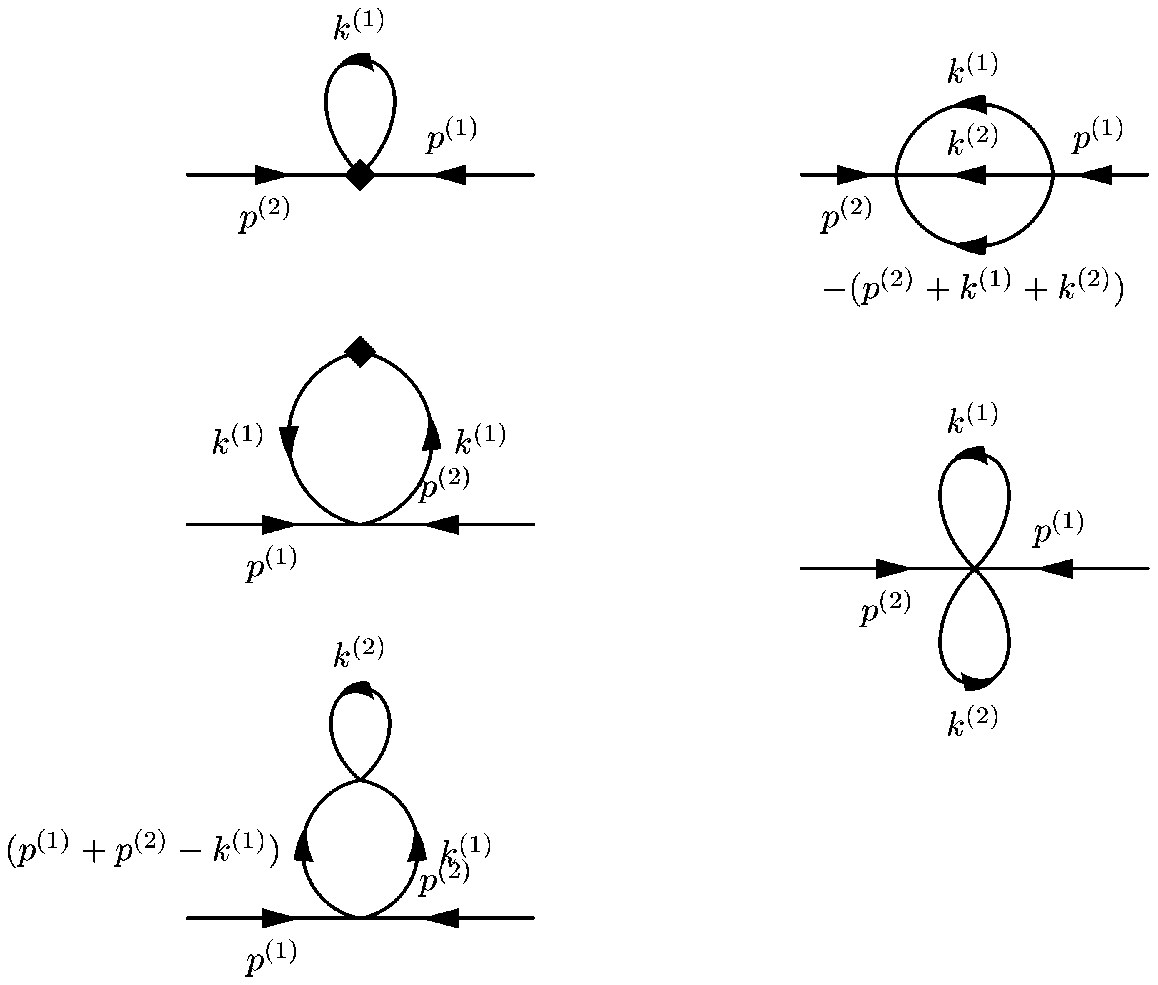}
\caption{(left) Fortran code generated by \textsc{MadeLine}. (right)
  FeynMF diagrams generated by \textsc{MadeLine}. }
\label{fig:diagrams}
\end{figure}

For this reason, a straightforward implementation of Wick's theorem is
implemented as a \textsc{Python} program called \textsc{MadeLine},
which creates the Feynman diagrams contributing to a specified
$n$-point function at a specified chiral order. The diagrams are
output both as \LaTeX/FeynMF figures and as Fortran functions using
the \textsc{chirpy/chirper} routines (Fig. \ref{fig:diagrams}).

Using the integration
routines provided by the \textsc{chirper} library, these functions can
be integrated numerically for an almost automated evaluation of the
corresponding lattice Feynman diagrams (where internal propagators
carry external on-shell momenta, some user intervention will be
necessary to assure the choice of the correct integration contour in
the complex plane).

\section{ Usage Overview }

\begin{figure}[h]
\begin{center}
\includegraphics[width=0.7\textwidth]{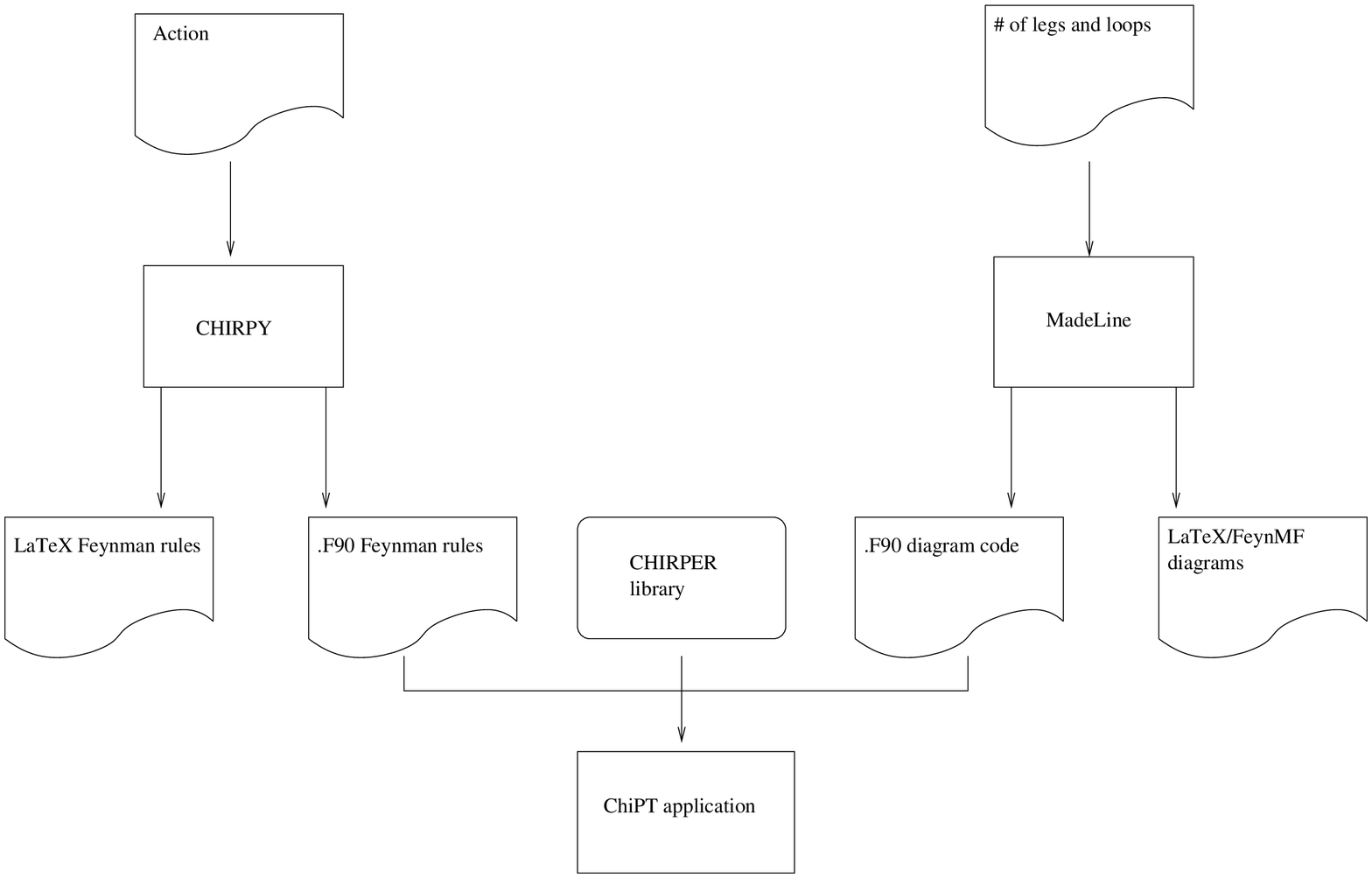} \\[2ex]
\end{center}
\caption{Schematic of usage of the tools}
\label{fig:usage}
\end{figure}

Fig. \ref{fig:usage}\/ shows a flow schematic of our tools:
The user specifies the action as an input to the \textsc{chirpy} main
script, which outputs the Feynman rules in the format chosen by the
user.

To compute a particular quantity, the user specifies the number
of legs and the loop order as input to the \textsc{MadeLine} script,
which outputs representations of the relevant Feynman diagrams in both
Fortran code and as graphics (where the \LaTeX/FeynMF output may have
to be edited by hand to achieve an esthetically pleasing result).

The generated code interfaces with the
\textsc{chirper} library and the Feynman rules generated by
\textsc{chirpy} in either compileable or machine-readable format. With
just a small amount of user-written code to bind these components
together, a fully functional application of lattice $\chi$PT can be
produced. Easy-to-use interfaces to initialisation and integration
routines are provided by \textsc{chirper}, making the user-written
code rather straightforward.

The structure of the interface between \textsc{chirpy} and
\textsc{MadeLine} allows to reuse the same Feynman rules for a
different set of diagrams and vice versa.

\section{ An Application to Finite-Volume Effects }

As an example of an application of this framework, we have repeated
the one-loop computation of the volume dependence of the pion mass
from \cite{borasoy}. The \textsc{chirper} software is so structured
that it is easy to evaluate the Feynman diagrams using a variety of
summation/integration methods, including VEGAS \cite{lepage}. This allows to
directly extract the difference between the finite-volume and
infinite-volume values of the diagrams (instead of using a large, but
finite volume to provide a value for the infinite-volume limit, as
done in \cite{borasoy}).

\begin{figure}[h]
\begin{center}
\includegraphics[width=7cm,angle=270,keepaspectratio=]{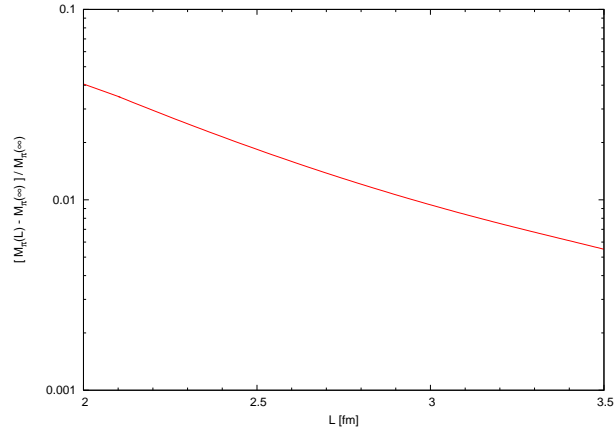}
\end{center}
\caption{Volume dependence of the pion mass at $M_\pi=140$ MeV}
\label{fig:voldep}
\end{figure}

The plot of Fig. \ref{fig:voldep}\/ shows an example of the results
which agree well with the earlier evaluation \cite{borasoy} that was
based on a hand-coded implementation of manually derived Feynman rules.

The great reduction of the workload in terms of human time by the
automated method makes a two-loop calculation appear viable, and a
first direct calculation of the two-loop volume dependence of the pion
mass is currently under way.

\section{ Conclusions }

Chiral Perturbation Theory can be formulated on the lattice, and
efficient methods from perturbative Lattice QCD can be translated to
the $\chi$PT case with only moderate effort. A software implementation
of an efficient algorithm  for generating perturbative expansions
exists. The Feynman rules and diagrams can be generated in different
formats suitable for use by either humans or computers. An almost
automated computation of lattice Feynman diagrams in lattice
regularised $\chi$PT is possible with these tools.
             
The automated procedure has successfully been used to repeat
finite-volume calculations previously done manually. New finite-volume
calculations at the two-loop level using the automated procedure are
currently under way.

\section*{ Acknowledgements }

GMvH wishes to thank R.R. Horgan and A.G. Hart for useful
discussions on automated methods in perturbative Lattice QCD.

This work was supported in part by the Natural Sciences
and Engineering  Research Council of Canada and by the Government of
Saskatchewan, and by the Deutsche Forschungsgemeinschaft.

\bibliographystyle{JHEP}
\bibliography{proceedings}

\end{document}